\documentclass[
aps,prd,
12pt,
nopreprintnumbers,
showpacs,
eqsecnum,
nofootinbib
]{revtex4-1}

\usepackage{graphicx}
\usepackage{amssymb}

\begin{document}

\title{Light fermion masses in partially deconstructed models}
\author{Nahomi Kan}\email[]{kan@gifu-nct.ac.jp}
\affiliation{National Institute of Technology, Gifu College,
Motosu-shi, Gifu 501-0495, Japan}
\author{Kiyoshi Shiraishi\footnote{Author to whom any
correspondence should be addressed.}}\email[]{shiraish@yamaguchi-u.ac.jp}
\affiliation{
Graduate School of Sciences and Technology for Innovation, Yamaguchi
University, Yamaguchi-shi, Yamaguchi 753--8512, Japan}
\author{Maki Takeuchi}\email[]{maki\_t@yamaguchi-u.ac.jp}
\affiliation{
Graduate School of Sciences and Technology for Innovation, Yamaguchi
University, Yamaguchi-shi, Yamaguchi 753--8512, Japan}
\date{\today}

\begin{abstract}
Considering a theory space consisting of a large number of five-dimensional Dirac
fermion field theories including background abelian gauge fields, we can construct
a theory similar to a continuous six-dimensional theory compactified with
two-dimensional manifolds with and without magnetic flux or orbifolds as extra
dimensions.  This method, called dimensional deconstruction, can be used to
construct a model with one-dimensional discrete space, which represents
general graph structures. In this paper, we propose the models with two extra
dimensions, which resemble two-dimensional tori, cylinders, and rectangular
regions, as continuum limits. We also try to build a model that mimics one with
the two-dimensional orbifold compactification.
\\ Keywords: deconstruction, orbifolds, compactification
\end{abstract}

\maketitle
\section{Introduction}
\label{introduction}
Higher-dimensional models are promising candidates for theories beyond the
standard model of elementary particles, and it is known that the geometry of the
extra dimensions is significantly related to the characteristics of the 
models \cite{KK}. Numerous studies have been conducted on compactification of
extra dimensions, while early work on compactification into manifolds discussed
the problem of chiral fermions. The problem was that four-dimensional chiral
fermions do not appear after compactification to a simple manifold.  In recent
years, however, it has been accepted that the chiral fermion problem can be solved
simply by considering a brane motivated by string theory and considering fermions
that exist only on it \cite{ADD}.

However, since the number of massless fermions is related to the number of
generations in the standard model, the emergence of four-dimensional chiral
fermions due to compactification of extra dimensions (without/with magnetic
flux) remains of interest. There are many models with extra dimensions,
but compactification onto orbifolds has been actively examined because it is
widely used in string theory, and it is  also known that
compactification onto orbifolds in field theory brings about the
appearance mechanism of four-dimensional chiral fermions.

On the other hand, a scheme of dimensional deconstruction was proposed about
twenty years ago as an idea for construction of analogs of compactified
theories \cite{ACG1,HPW,Lane,HL}. Its original motivation was to construct 
Higgsless models \cite{ACG2,Schmaltz} as alternatives to the higher-dimensional
gauge-Higgs unified theory \cite{Hosotani1,Hosotani2,HIL}. The idea of dimensional
deconstruction is that by means of connecting multiple copies of a
four-dimensional field theory with a link field similar to a gauge field,
 one can present a theory similar to a higher-dimensional model. In
models based on this idea, the degrees of freedom of the link fields play the role
of the Higgs fields. In addition, in the case of the special type of scalar
theory considered in the paper \cite{CKS}, it was shown that a non-trivial link
field produce the same effect as a magnetic flux in extra dimensions.

In the present paper, we propose a new model that can be interpreted as a hybrid
of compactification and deconstruction of extra dimensions. Although only simple
examples will be given here, future study may address the following
prospects: One is to see, if there are small differences between the
``manifold/orbifold mimicker'' which will be described in the following sections
and the corresponding continuous model,
it will serve as an alternative to confirm the degree of certainty of the
model with minimum requirements from observable physical quantities. Another is
that, since the simple structure of the deconstruction theory can be replaced with
the structure of a general graph, so further extensions of the model can be
discussed in the future. Other additional issues will be discussed in the final
section. Although the model treated in this paper is a toy model as a simple
example,  it is possible to discuss models with which we can perform one-loop
quantum calculations including vacuum energy, because the masses of
all excited states can be obtained, in principle, in our model. For this reason, 
further analysis on quantum effects can also be expected in further extended
models.

The organization of this paper is as follows. In the next Section 2, we propose
the general form of the toy model with graph structure considered below. In
Section 3, we review the compactification of the five-dimensional Dirac field
theory, which is a component of the model, and prepare for the construction of
the entire model. The fermion mass matrix and its spectrum, when a cycle graph is
adopted as the graph, will be discussed in Section 4. This is a model in which the
extra dimensions are a two-dimensional torus or a cylinder in the continuum
limit. On the other hand, in Section 5, we consider the fermion mass matrix
and its spectrum, when a path graph is used. In this case, it is a model that
gives extra dimensions of a cylinder or a rectangle in the continuum limit. We
try to construct a model in which the continuum limit looks like an orbifold
$T^2/Z_2$ in the last part of the section. The final section provides a summary
and future prospects. We use the metric convention
$(-+\cdots+)$.

\section{The model}
\label{sec2}

Let $G(\mathcal{V}, \mathcal{E})$ be a graph with vertex set $\mathcal{V}$ and
edge set $\mathcal{E}$.%
\footnote{For the (spectral) graph theory, see
\cite{Mohar1,Mohar2,Mohar3,Merris}.} We here consider a directed graph. An
oriented edge is represented by $e=[u, v]$ ($u, v\in\mathcal{V}(G)$). One of
these vertices, say $u$, is the open of the edge, denoted by $o(e)$, while the
other vertex is the terminus of the edge, denoted by $t(e)$.

We consider two sets of five-dimensional Dirac fields $\Psi_v$ and
$\tilde{\Psi}_e$, which are assigned to vertices (sites) and edges (links),
respectively. The model that imitates the six-dimensional theory considered here
is described by the Lagrangian density
\begin{equation}
\mathcal{L}=\sum_{v\in\mathcal{V}(G)}\mathcal{L}_K[\Psi_v; A_{M,v}]+
\sum_{e\in\mathcal{E}(G)}\mathcal{L}_K[\tilde{\Psi}_e;A_{M,e}]+
\mathcal{L}_M[\Psi_v;\tilde{\Psi}_e]\,,
\end{equation}
where
\begin{equation}
\mathcal{L}_K[\Psi;A_M]=i\overline{\Psi}\Gamma^MD_M\Psi\,,
\end{equation}
with
\begin{equation}
D_M=\partial_M+iA_M\,,
\end{equation}
and
\begin{equation}
\mathcal{L}_M[\Psi_v,\tilde{\Psi}_e]=-m
\sum_{e\in\mathcal{E}
(G)}\left[\overline{\tilde{\Psi}}_e(\Psi_{t(e)}-\Psi_{o(e)})+h.c.\right]\,.
\end{equation}
Here, $A_M$ $(M=0,1,2,3,5)$ denotes a $U(1)$ gauge field and $m$ is a constant
with dimension of mass.
The reason why we consider two types of Dirac fields is that, in six dimensions,
fermions have twice as many degrees of freedom as in five dimensions.
Under the $U(1)$ gauge transformation, it follows that all fields undergo a
common local phase transformation. This is due to the term $\mathcal{L}_M$.

The gamma matrices for five dimensions satisfies $\{\Gamma^M,
\Gamma^N\}=-2\eta^{MN}$  and
$\overline{\Psi}\equiv \Psi^\dagger\Gamma^0$.
The five dimensional gamma matrices can be expressed by the usual four dimensional
gamma matrices $\gamma^\mu$, as $\Gamma^\mu=\gamma^\mu$ $(\mu=0,1,2,3)$ and 
$\Gamma^5=i\gamma^5$, where $(\gamma^5)^2=1$ is satisfied.

The term $\mathcal{L}_M$ can be rewritten by
\begin{equation}
\mathcal{L}_M[\Psi_v,\tilde{\Psi}_e]=m\sum_{v\in\mathcal{V}(G)}\sum_{e\in
\mathcal{E}(G)}\left(\overline{\tilde{\Psi}}_e E^T_{ev}(G)\Psi_v+h.c.\right)\,,
\end{equation}
where $E(G)$ is the incidence matrix for a directed graph $G$ defined by
\begin{equation}
E_{ve}(G)=\left\{\begin{array}{cc}
1 & \mbox{if~} v=o(e) \\
-1 & \mbox{if~} v=t(e) \\
0 & \mbox{otherwise}
\end{array}
\right.\,.
\end{equation}

\begin{figure}[ht]
\centering
\includegraphics
{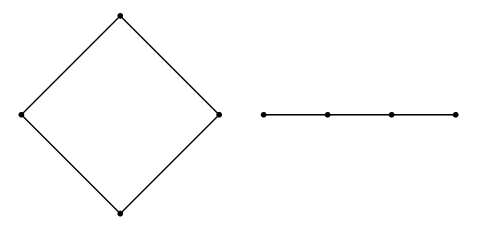}
\caption{Left: the cycle graph $C_4$. Right: the path graph $P_4$.
Each suffix indicates the number of vertices in the graph.}
\label{fig2_1}
\end{figure}

In this paper, only $G=C_N$ and $G=P_{N+1}$ are used as graphs.
The cycle graph $C_N$ has a structure in which $N$ vertices are connected by $N$
edges like a ring (left of Fig.~\ref{fig2_1}). For example, the incidence matrix
of
$C_4$ is expressed as
\begin{equation}
E(C_4)=\left[\begin{array}{rrrr}
1 & 0 & 0 & -1\\
-1 & 1 & 0 & 0\\
0 & -1 & 1 & 0 \\
0 & 0 & -1 & 1
\end{array}
\right]\,,
\end{equation}
where sequence numbers of vertices and edges are supposed to be
appropriately assigned.

A path graph $P_{N}$ is a graph in which $N$ vertices are connected by $N-1$ edges
like a single path (right of Fig.~\ref{fig2_1}). For example, the incidence matrix
of
$P_4$ is
\begin{equation}
E(P_4)=\left[\begin{array}{rrrr}
1 & 0 & 0 \\
-1 & 1 & 0 \\
0 & -1 & 1  \\
0 & 0 & -1 
\end{array}
\right]\,,
\end{equation}
where sequence numbers of vertices and edges are supposed to be
appropriately assigned.

Incidentally, the graph Laplacian \cite{Mohar1,Mohar2,Mohar3,Merris} can be
defined as
\begin{equation}
\Delta(G)=E(G) E^T(G)\,.
\end{equation}
For example, we find%
\footnote{Note that $\Delta(C_N)=E(C_N) E^T(C_N)=E^T(C_N) E(C_N)$ holds true only
for cycle graphs.}
\begin{equation}
\Delta(C_4)=\left[\begin{array}{rrrr}
2 & -1& 0 & -1\\
-1 & 2 & -1 & 0\\
0 & -1 &2 & -1 \\
-1 & 0 & -1 & 2
\end{array}
\right]\,,
\end{equation}
and
\begin{equation}
\Delta(P_4)=\left[\begin{array}{rrrrr}
1 & -1 & 0 & 0\\
-1 & 2 & -1& 0 \\
0 & -1 &2 &-1  \\
0 & 0 & -1 &1
\end{array}
\right]\,.
\end{equation}

Although it can be intuitively anticipated that $C_N$ is a discretization of $S^1$
(a circle)
and $P_N$ is a discretization of $S^1/Z_2$ (a line segment), the continuum limit
of the current model will actually be clarified in the following sections.

To obtain a four-dimensional theory, it is necessary to compactify one dimension
of each five-dimensional theory. For the fifth dimension, we consider the
manifold $S^1$ or the orbifold $S^1/Z_2$. The compactification is reviewed in the
next section.
In the following sections, the four-dimensional components of the gauge field are
assumed to be zero, as a trivial background field.

\section{compactification of the five-dimensional theory}
\label{sec3}

In this section, we consider compactification of the five-dimensional fermion
theory as a component in ``theory space''. The theme of this paper is to
construct a higher-dimensional model mimicker (imitator) with two extra dimensions
by stitching these $N$ copies together using a graph structure.

\subsection{compactification on $S^1$}

Let $y$ be the coordinate of the fifth dimension ($y=x^5$).
Since we set a periodic boundary condition for $y$,
\begin{equation}
\Psi(x,y+2\pi R)=\Psi(x,y)\,,
\end{equation}
where $x$ denotes the four-dimensional coordinates, $x^\mu$ $(\mu=0, 1, 2, 3)$,
and $R$ is a constant with length dimension. Then, the Dirac field can be
expanded in a Fourier series with respect to
$y$ as
\begin{equation}
\Psi(x,y)=\frac{1}{\sqrt{2\pi
R}}\sum_{n=-\infty}^\infty\psi_n(x)e^{i\frac{n}{R}y}\,,
\end{equation}
where $\psi_n(x)$ are regarded as four-dimensional Dirac fields.

We here assume that there is a zero mode of the gauge field $A_y$ in the
fifth direction, which can have a constant vacuum expectation value,
$\langle A_y\rangle$. Thus the kinetic part of the Lagrangian reads
\begin{equation}
\mathcal{L}_K[\Psi;A_y]=i\overline{\Psi}\Gamma^MD_M\Psi=i\overline{\Psi}
\left[\Gamma^\mu\partial_\mu+\Gamma^5(\partial_y+i\langle
A_y\rangle)\right]\Psi\,.
\end{equation}

Integrating this kinetic part of the Lagrangian over the extra coordinate $y$,
we obtain the Lagrangian for four-dimensional fields,
\begin{eqnarray}
& &\int_0^{2\pi R} \mathcal{L}_K[\Psi;A_y] dy=\sum_{n=-\infty}^{\infty}
\left[i\overline{\psi}_n\gamma^\mu\partial_\mu\psi_n-
i\frac{n+\delta}{R}\overline{\psi}_n\gamma^5\psi_n\right]\nonumber \\
&=&\sum_{n=-\infty}^{\infty}
\left[i\overline{\psi}_{Ln}\gamma^\mu\partial_\mu\psi_{Ln}+
i\overline{\psi}_{Rn}\gamma^\mu\partial_\mu\psi_{Rn}+
i\frac{n+\delta}{R}\left(\overline{\psi}_{Rn}\psi_{Ln}-
\overline{\psi}_{Ln}\psi_{Rn}\right)\right]\,,
\end{eqnarray}
where $\delta=R\langle A_y\rangle$.
In this expression, the decomposition $\psi=\psi_L+\psi_R$ is done according to
$\psi_L=\frac{1-\gamma^5}{2}\psi$ and $\psi_R=\frac{1+\gamma^5}{2}\psi$.

\subsection{compactification on $S^1/Z_2$}

Next, we consider compactification on $S^1/Z_2$.
It is the simplest scheme producing a four-dimensional chiral fermion from a
five-dimensional Dirac field \cite{PQ}.
On $S^1/Z_2$, in addition to the periodic condition $\Psi(x,y+2\pi R)=\Psi(x,y)$,
we require the conditions
\begin{equation}
\Psi(x, -y)=P_0\gamma^5\Psi(x, y)\,,\quad
\Psi(x, \pi R-y)=P_1\gamma^5\Psi(x, \pi R+y)
\quad (P_{0,1}=\pm 1)\,,
\end{equation}
which realize two fixed points of the orbifold $S^1/Z_2$, $y_0=0$ and $y_1=\pi R$.
If we choose $P_0=P_1=-1$, we find
\begin{eqnarray}
& &\Psi_L(x, -y)=+\Psi_L(x,y)\,,\quad
\Psi_L(x, \pi R-y)=+\Psi_L(x,\pi R+y)\,,\\
& &\Psi_R(x, -y)=-\Psi_R(x,y)\,,\quad
\Psi_R(x, \pi R-y)=-\Psi_R(x,\pi R+y)\,,
\end{eqnarray}
and thus, the fermion field can be expanded as
\begin{eqnarray}
\Psi_L(x, y)&=&\frac{1}{\sqrt{2\pi R}}\psi_{L0}(x)+\sqrt{\frac{1}{\pi R}}
\sum_{n=1}^\infty\psi_{Ln}(x)\cos\frac{ny}{R}\,,\\
\Psi_R(x, y)&=&i\sqrt{\frac{1}{\pi R}}
\sum_{n=1}^\infty\psi_{Rn}(x)\sin\frac{ny}{R}\,.
\end{eqnarray}
If we choose $P_0=P_1=+1$, the left and right handed fermions are swapped
from the above expressions.
Note that  a zero mode of the gauge field may exist
when an unusual, exotic condition in the fifth direction is assumed,
but it is not considered here, and so we suppose $\langle A_y\rangle=0$
for the $S^1/Z_2$ compactification throughout this paper.

The kinetic part of the Lagrangian can be expressed in terms of the
four-dimensional fields, after integration over $y$, as
\begin{eqnarray}
& &\int_0^{2\pi R} \mathcal{L}_K[\Psi;0] dy \nonumber\\
&=&
i\overline{\psi}_{L0}\gamma^\mu\partial_\mu\psi_{L0}+\sum_{n=1}^{\infty}
\left[i\overline{\psi}_{Ln}\gamma^\mu\partial_\mu\psi_{Ln}+
i\overline{\psi}_{Rn}\gamma^\mu\partial_\mu\psi_{Rn}+i
\frac{n}{R}\left(\overline{\psi}_{Rn}\psi_{Ln}-
\overline{\psi}_{Ln}\psi_{Rn}\right)\right]\,.
\end{eqnarray}

In the following sections, the model will be described when a graph and
compactification are specified. In the following sections, for instance, a model
that employs
$S^1$ compactification and a $C_N$ graph will be abbreviated as the $S^1\otimes
C_N$ model.

\section{deconstruction with $C_N$}
\label{sec4}

\begin{figure}[ht]
\centering
\includegraphics[width=5cm]
{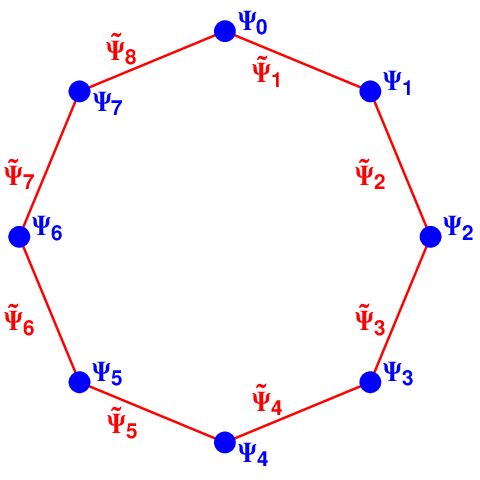}
\caption{A cycle graph $C_8$ as a Moose diagram of the model.}
\label{fig4_1}
\end{figure}

In this section we adopt a cycle graph $C_N$ (Fig.~\ref{fig4_1}) in our model.
We label the fields as
$\Psi_p(x,y)$ $(p=0, 1, \dots, N-1)$, and $\tilde{\Psi}_p(x,y)$  $(p=1, 2,
\dots, N)$.
Then, we find
\begin{equation}
-m\sum_{e\in
\mathcal{E}(C_N)}
\left[\overline{\tilde{\Psi}}_e(\Psi_{t(e)}-\Psi_{o(e)})+h.c.\right]
=-m\sum_{p=1}^N\left[\overline{\tilde{\Psi}}_p(\Psi_{p}-\Psi_{p-1})+h.c.\right]\,,
\end{equation}
where, in the sum, the identification $\Psi_N=\Psi_0$ is required.

\subsection{$S^1\otimes C_N$}
\begin{figure}[ht]
\centering
\includegraphics[width=5cm]
{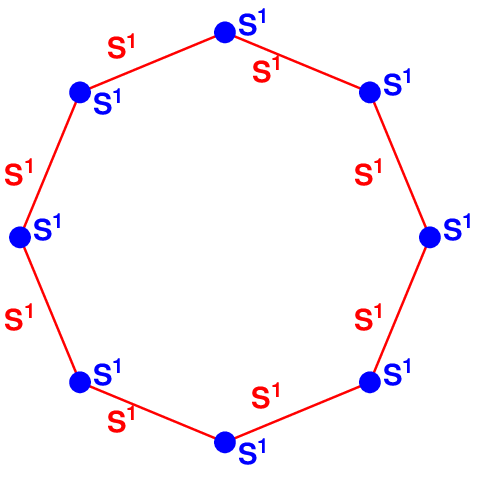}
\caption{Graphical representation of the $S^1\otimes C_N$ model ($N=8$).}
\label{fig4_2}
\end{figure}

We assume that $\tilde{\Psi}_p$ are compactified on a common radius $R$ of $S^1$,
as well as ${\Psi}_p$ (Fig.~\ref{fig4_2}).
Then, we simply obtain
\begin{eqnarray}
& &\int_0^{2\pi
R}\mathcal{L}_M[\Psi_p,
\tilde{\Psi}_p]dy=-m\sum_{n=-\infty}^\infty\sum_{p=1}^N\left(\overline{\tilde{\psi}}_{p,n}
(\psi_{p,n}-\psi_{p-1,n})+h.c.\right)\nonumber \\
& &=-m\sum_{n=-\infty}^\infty\sum_{p=1}^N\left(\overline{\tilde{\psi}}_{R,p,n}
(\psi_{L,p,n}-\psi_{L,p-1,n})+\overline{\tilde{\psi}}_{L,p,n}
(\psi_{R,p,n}-\psi_{R,p-1,n})+h.c.\right)\,.
\end{eqnarray}

Considering the mass terms that come out of compactification together, the total
mass term of component fields can be written as follows:
\begin{equation}
-\sum_{n=-\infty}^\infty\begin{array}{rr}
\Bigl[\overline{\psi}_n & \overline{\tilde{\psi}}_n\Bigr]_R\\
&
\end{array}
\left[\begin{array}{rr}
M_n & {M'}^T_n\\
{M'_n} & \tilde{M}_n
\end{array}
\right]
\left[\begin{array}{r}
\psi_n \\
\tilde{\psi}_n
\end{array}
\right]_L+h.c.\,,
\label{massterm}
\end{equation}
where $\psi_n=(\psi_{0,n},\dots\psi_{N-1,n})^T$ and
$\tilde{\psi}_n=(\tilde{\psi}_{1,n},\dots\tilde{\psi}_{N,n})^T$ with
\begin{equation}
M_n=\left[\begin{array}{rrrrr}
-i\frac{n+\delta_0}{R} &  & & &\\
 & -i\frac{n+\delta_1}{R} &  & &   \\
 &  &  & \ddots &   \\
 &  &  & &  -i\frac{n+\delta_{N-1}}{R}
\end{array}
\right]\,,\quad
\tilde{M}_n=\left[\begin{array}{rrrrr}
-i\frac{n+\tilde{\delta}_1}{R} &  & & &\\
 & -i\frac{n+\tilde{\delta}_2}{R} &  & &   \\
 &  &  & \ddots &   \\
 &  &  & &  -i\frac{n+\tilde{\delta}_{N}}{R}
\end{array}
\right]\,,
\end{equation}
and
\begin{equation}
{M'}_n=m\left[\begin{array}{rrrrr}
-1 & 1 & 0  &\cdots & 0\\
0  & -1 & ~1  &\cdots  &0  \\
\vdots & \vdots & \vdots & \ddots & \vdots  \\
1 & 0 & 0 & \cdots  & -1
\end{array}
\right]=-m E^T(C_N)\,,
\end{equation}
and the contributions of the background gauge fields come into
play as follows: $\delta_p\equiv R\langle (A_y)_{v=p}\rangle$ and
$\tilde{\delta}_p\equiv R\langle (\tilde{A}_y)_{e=p}\rangle$.

From the $n$th mass matrix 
\begin{equation}
\mathcal{M}_n=\left[\begin{array}{rr}
M_n & {M'}^T_n\\
{M'_n} & \tilde{M}_n
\end{array}
\right]\,,
\end{equation}
we obtain the $n$th mass-squared matrix
\begin{equation}
\mathcal{M}^2_n\equiv\mathcal{M}^\dagger_n\mathcal{M}_n=\left[\begin{array}{rr}
M_n^\dagger M_n+{M'}^T_nM'_n& M_n^\dagger {M'}^T_n+{M'}^T_n\tilde{M}_n\\
 {M'}_nM_n+\tilde{M}^\dagger_n{M'}_n & \tilde{M}_n^\dagger
\tilde{M}_n+{M'}_n{M'}^T_n
\end{array}
\right]\,.
\end{equation}

If $\delta_0=\delta_1=\cdots=\delta_{N-1}=
\tilde{\delta}_1=\tilde{\delta}_2=\cdots=\tilde{\delta}_{N}\equiv\delta$ are
assumed, we find
\begin{equation}
\mathcal{M}^2_n=\left[\begin{array}{cc}
\mathbf{M}^2_n & 0\\
0 & \mathbf{M}^2_n
\end{array}
\right]\,,
\end{equation}
where
\begin{equation}
\mathbf{M}^2_n=\frac{(n+\delta)^2}{R^2}+m^2\Delta(C_N)\,.
\end{equation}

Therefore, the eigenvalues of the mass-squared of fermions are \cite{Mohar3}
\begin{equation}
\frac{(n+\delta)^2}{R^2}+4m^2\sin^2\frac{\pi k}{N}\quad
(n=\cdots, -1, 0, 1, \cdots, k=0, 1,\cdots, N-1)\,.
\end{equation}
Two (for $\psi$ and $\tilde{\psi}$) Dirac zero modes are found for $n=0$ and
$k=0$, when
$\delta=0$. 

If we further replace
$m\rightarrow\frac{N}{2\pi\tilde{R}}$ and taking the limit of
$N\rightarrow\infty$, the eigenvalues become
\begin{equation}
\frac{(n+\delta)^2}{R^2}+\frac{k^2}{\tilde{R}^2}\,.
\end{equation}

\begin{figure}[ht]
\centering
\includegraphics[width=6cm]
{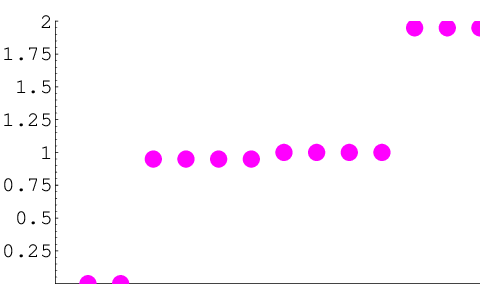}
\caption{The first several eigenvalues of the mass-squared matrix of
the $S^1\otimes C_N$ model with
$m=\frac{N}{2\pi\tilde{R}}$, $R=\tilde{R}=1$, $\delta=0$, and $N=8$.}
\label{fig4_3}
\end{figure}

We thus recognized that the continuum limit of the model in this section
is a theory compactified on a two-torus $T^2$ (topologically equivalent to a
surface of a doughnut). Fig.~\ref{fig4_3} shows the small eigenvalues of the
mass-squared matrix for
$\delta=0$,
$m=\frac{N}{2\pi\tilde{R}}$, $R=\tilde{R}=1$, and $N=8$.

\subsection{$S^1\otimes C_N$ with flux}

In a continuous theory with torus compactification, a homogeneous flux on the
two-torus, whose coordinates $(x, y)$ satisfy $x\sim x+2\pi\tilde{R}$ and $y\sim
y+2\pi R$, is described by the gauge field
\begin{equation}
A_y=fx\quad (f \mbox{~is a constant})\,,
\end{equation}
and then, the constant magnetic flux density is given by
\begin{equation}
F_{xy}=\partial_xA_y-\partial_yA_x=f\,.
\end{equation}
If the torus is expressed by $x\sim x+2\pi\tilde{R}$,
$A_y$ and $A_y+2\pi\tilde{R}f$ must be equivalent.
On the other hand, the degree of freedom of the gauge transformation
\begin{equation}
\Psi\rightarrow e^{i\xi}\Psi\, ,\quad A_y\rightarrow A_y+\partial_y\xi
\end{equation}
implies the identification of the gauge field on the torus described by $y\sim
y+2\pi R$, that is,
\begin{equation}
A_y\sim A_y+\frac{q}{R}\quad (q\mbox{: integer}).
\end{equation}
Therefore, consistency on the torus requires $f=\frac{q}{2\pi R\tilde{R}}$ and
thus 
\begin{equation}
A_y=\frac{q}{2\pi R\tilde{R}}x\quad (q \mbox{~is a constant})\,.
\label{contF}
\end{equation}

Now, in our $S^1\otimes C_N$ model, we think of $C_N$ as a discretized circle
with radius $\tilde{R}$, it can be interpreted as equating $x/(2\pi\tilde{R})$
in the continuous theory,
with an integer multiple of $1/N$, so the background gauge
fields on vertices $\delta_p=R\langle A_y
\rangle$ can be read as
\begin{equation}
\delta_p=\frac{qp}{N}\quad(p=0,1,\dots, N-1)\,,
\label{f1}
\end{equation}
for an integer $q$. Furthermore, if we set simply 
\begin{equation}
\tilde{\delta}_p=\frac{qp}{N}\quad(p=1,\dots, N)\,,
\label{f2}
\end{equation}
for the background gauge fields on edges.
These sets of configurations yield a model whose continuum limit gives the torus
compactification with constant flux. 

\begin{figure}[ht]
\centering
\includegraphics[width=6cm]
{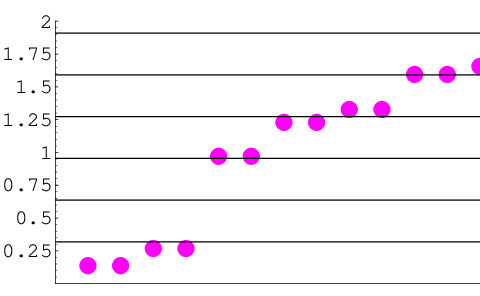}
\caption{The first several eigenvalues of the mass-squared matrix in the case that
$\delta_p$ and $\tilde{\delta}_p$ are given by (\ref{f1}) and (\ref{f2}) with
$q=1$, respectively. The other settings are,
$m=\frac{N}{2\pi\tilde{R}}$, $R=\tilde{R}=1$, and $N=8$.
The horizontal lines represent integer multiples of $\frac{1}{\pi}$.}
\label{fig4_4}
\end{figure}
\begin{figure}[ht]
\centering
\includegraphics[width=6cm]
{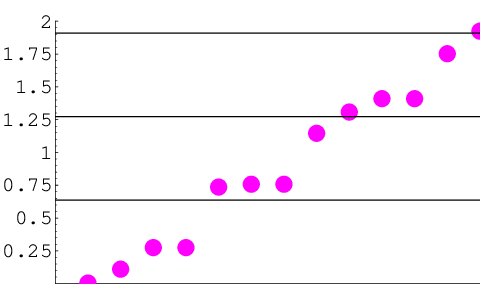}
\caption{The first several eigenvalues of the mass-squared matrix in the case that
$\delta_p$ and $\tilde{\delta}_p$ are given by (\ref{f1}) and (\ref{f2}) with
$q=2$, respectively. The other settings are,
$m=\frac{N}{2\pi\tilde{R}}$, $R=\tilde{R}=1$, and $N=8$.
The horizontal lines represent integer multiples of $\frac{2}{\pi}$.}
\label{fig4_5}
\end{figure}
\begin{figure}[ht]
\centering
\includegraphics[width=6cm]
{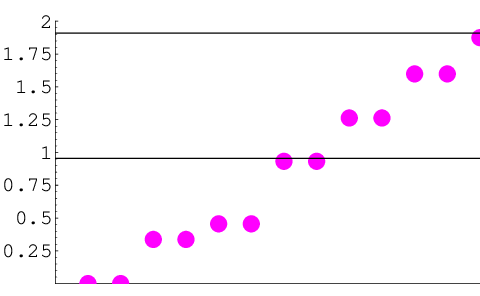}
\caption{The first several eigenvalues of the mass-squared matrix in the case that
$\delta_p$ and $\tilde{\delta}_p$ are given by (\ref{f1}) and (\ref{f2}) with
$q=3$, respectively. The other settings are,
$m=\frac{N}{2\pi\tilde{R}}$, $R=\tilde{R}=1$, and $N=8$.
The horizontal lines represent integer multiples of $\frac{3}{\pi}$.}
\label{fig4_6}
\end{figure}

The first several eigenvalues of the mass-squared matrix for
$m=\frac{N}{2\pi\tilde{R}}$, $R=\tilde{R}=1$, and $N=8$ are shown in
Fig.~\ref{fig4_4}--\ref{fig4_6}. Fig.~\ref{fig4_4} shows the case with $q=1$,
Fig.~\ref{fig4_5} shows the case with $q=2$, and Fig.~\ref{fig4_6} shows the case
with $q=3$.

In the continuous theory with the background gauge field (\ref{contF}), the
eigenvalues are expressed by \cite{DS,DSW}
\begin{equation}
\frac{q}{\pi R\tilde{R}}k\quad(k=0,1,2,\dots)\,,
\end{equation}
and the degeneracy of the zero modes is $2 q$, because two species $\psi$
and
$\tilde{\psi}$ exist in our model.

In our result shown in Fig.~\ref{fig4_4}--\ref{fig4_6}, the first relatively large
mass gap appears approximately at the expected location in the continuum limit
(however, in the case of $q=1$, it is not so clear after that).
The reason that each eigenvalue that should be degenerate in the continuum limit
is not so degenerate for finite $N$ is due to the large anisotropy in the extra
two (continuous and discrete) directions, which is unavoidable.
For example, it is known from previous study that degeneracy is guaranteed to some
extent in a lattice model that is discrete in both directions \cite{TT}.

\subsection{$S^1/Z_2\otimes C_N$}

\begin{figure}[ht]
\centering
\includegraphics[width=5cm]
{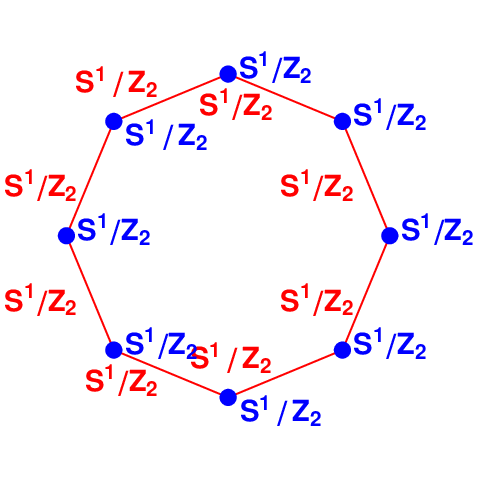}
\caption{Graphical representation of the $S^1/Z_2\otimes C_N$ model ($N=8$).}
\label{fig4_7}
\end{figure}

In this case, we suppose that the fermions on the edges satisfy the reverse
condition $P_0=P_1=+1$, while the fermions on the vertices satisfy the condition
$P_0=P_1=-1$. Otherwise, we find that $\mathcal{L}_M\equiv 0$ if all fermions are
governed by the same condition,and it  results in a trivial set of
$N$ individual five-dimensional theories.%

Now, we obtain, for the model consisting of $S^1/Z_2$ compactification with $C_N$,
\begin{eqnarray}
& &\int_0^{2\pi
R}\mathcal{L}_M[\Psi_p,
\tilde{\Psi}_p]dy=-m\sum_{p=1}^N\left[\overline{\tilde{\psi}}_{R,p,0}
(\psi_{L,p,0}-\psi_{L,p-1,0})+h.c.\right]\nonumber \\
& &-m\sum_{n=1}^\infty\sum_{p=1}^N\left[\overline{\tilde{\psi}}_{R,p,n}
(\psi_{L,p,n}-\psi_{L,p-1,n})+
\overline{\tilde{\psi}}_{L,p,n}
(\psi_{R,p,n}-\psi_{R,p-1,n})+h.c.\right]\,.
\end{eqnarray}

The eigenvalues of the mass squared of fermions in this case are
\begin{equation}
\frac{n^2}{R^2}+4m^2\sin^2\frac{\pi k}{N}\quad
(n=0, 1, \cdots, k=0, 1,\cdots, N-1)\,.
\end{equation}
One zero mode is expressed by $\psi_{L,0,0}=\psi_{L,1,0}=\cdots=\psi_{L,N-1,0}$,
and one mode is expressed by
$\tilde{\psi}_{R,1,0}=\tilde{\psi}_{R,2,0}=\cdots=\tilde{\psi}_{R,N,0}$. 
These two modes are equivalent to one Dirac zero mode. If we
replace
$m\rightarrow\frac{N}{2\pi\tilde{R}}$ and taking the limit of
$N\rightarrow\infty$, the eigenvalues become
\begin{equation}
\frac{n^2}{R^2}+\frac{k^2}{\tilde{R}^2}\,.
\end{equation}
Further, if $\tilde{R}=R$, six (Dirac) modes belong to the second-smallest
eigenvalue.
We thus recognized that the continuum limit of the model in this section
is a theory compactified on a cylinder.

\section{deconstruction with $P_{N+1}$}
\label{sec5}

\begin{figure}[ht]
\centering
\includegraphics
{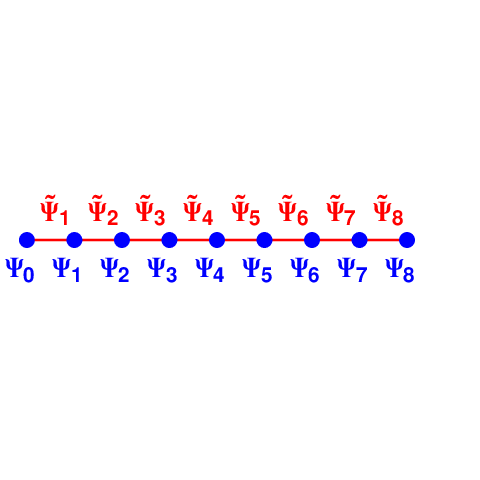}
\caption{A path graph $P_9$ as a Moose diagram of the model.}
\label{fig5_1}
\end{figure}

In this section we adopt a path graph $P_{N+1}$ in the primary model.
We label the fields as
$\Psi_p(x,y)$ $(p=0, 1, \dots, N)$, and $\tilde{\Psi}_p(x,y)$  $(p=1, 2,
\dots, N)$ (Fig.~\ref{fig5_1}).
Then, we find
\begin{equation}
-m\sum_{e\in
\mathcal{E}(P_{N+1})}
\left[\overline{\tilde{\Psi}}_e(\Psi_{t(e)}-\Psi_{o(e)})+h.c.\right]
=-m\sum_{p=1}^N\left[\overline{\tilde{\Psi}}_p(\Psi_{p}-\Psi_{p-1})+h.c.\right]\,,
\end{equation}

\subsection{$S^1\otimes P_{N+1}$}

\begin{figure}[ht]
\centering
\includegraphics
{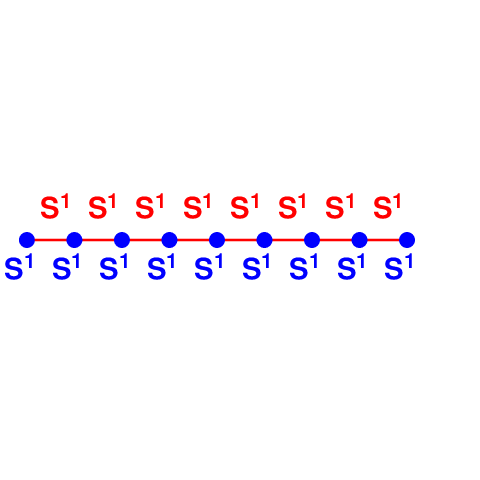}
\caption{Graphical representation of the $S^1\otimes P_{N+1}$ model ($N=8$).}
\label{fig5_2}
\end{figure}

We assume that $\tilde{\Psi}_p$ are compactified on a common radius $R$ of
$S^1$, as well as ${\Psi}_p$, as before (Fig.~\ref{fig5_2}).
Then, we simply obtain
\begin{eqnarray}
& &\int_0^{2\pi
R}\mathcal{L}_M[\Psi_p,
\tilde{\Psi}_p]dy=-m\sum_{n=-\infty}^\infty\sum_{p=1}^N\left[\overline{\tilde{\psi}}_{p,n}
(\psi_{p,n}-\psi_{p-1,n})+h.c.\right]\nonumber \\
& &=-m\sum_{n=-\infty}^\infty\sum_{p=1}^N\left[\overline{\tilde{\psi}}_{R,p,n}
(\psi_{L,p,n}-\psi_{L,p-1,n})+\overline{\tilde{\psi}}_{L,p,n}
(\psi_{R,p,n}-\psi_{R,p-1,n})+h.c.\right]\,.
\end{eqnarray}
This apparently describes non-chiral theory.
The fermion mass term can be written similarly to (\ref{massterm}) in the
$S^1\otimes C_{N}$ model. In the present case, the $n$th mass matrix is given by
\begin{equation}
\mathcal{M}_n=\left[\begin{array}{rr}
M_n & {M'}^T_n\\
{M'_n} & \tilde{M}_n
\end{array}
\right]\,,
\end{equation}
where
\begin{equation}
M_n=\left[\begin{array}{rrrrr}
-i\frac{n+\delta_0}{R} &  & & &\\
 & -i\frac{n+\delta_1}{R} &  & &   \\
 &  &  & \ddots &   \\
 &  &  & &  -i\frac{n+\delta_{N}}{R}
\end{array}
\right]\,,\quad
\tilde{M}_n=\left[\begin{array}{rrrrr}
-i\frac{n+\tilde{\delta}_1}{R} &  & & &\\
 & -i\frac{n+\tilde{\delta}_2}{R} &  & &   \\
 &  &  & \ddots &   \\
 &  &  & &  -i\frac{n+\tilde{\delta}_{N}}{R}
\end{array}
\right]\,,
\end{equation}
and
\begin{equation}
{M'}_n=m\left[\begin{array}{rrrrrr}
-1 & 1 & 0  &\cdots & 0 & 0\\
0  & -1 & ~1  &\cdots  &0 &0 \\
\vdots & \vdots & \vdots & \ddots & \vdots& \vdots  \\
0 & 0 & 0 & \cdots  & -1& ~1
\end{array}
\right]=-m E^T(P_{N+1})\,.
\end{equation}

Further, if $\delta_0=\delta_1=\cdots=\delta_{N-1}=
\tilde{\delta}_1=\tilde{\delta}_2=\cdots=\tilde{\delta}_{N}\equiv\delta$ is
assumed, we find
\begin{equation}
\mathcal{M}^2_n=\left[\begin{array}{cc}
\mathbf{M}^2_n & 0\\
0 & \tilde\mathbf{M}^2_n
\end{array}
\right]\,,
\end{equation}
where
\begin{equation}
\mathbf{M}^2_n=\frac{(n+\delta)^2}{R^2}+m^2\Delta(P_{N+1})\,,\quad
\tilde\mathbf{M}^2_n=\frac{(n+\delta)^2}{R^2}+m^2L(P_{N+1})\,.
\end{equation}
Here, $L(G)\equiv E^T(G)E(G)$ is the Laplacian of the line graph of $G$
\cite{Mohar1,Merris}.

The eigenvalues of $\mathbf{M}^2_n$ are \cite{Mohar1}
\begin{equation}
\frac{(n+\delta)^2}{R^2}+4m^2\sin^2\frac{\pi k}{2(N+1)}\quad
(n=\cdots, -1, 0, 1, \cdots, k=0, 1,\cdots, N)\,,
\end{equation}
while the eigenvalues of $\tilde\mathbf{M}^2_n$ are%
\footnote{Note the fact that the eigenvalues of matrices $AB$ and $BA$ are
identical except for zero eigenvalues.}
\begin{equation}
\frac{(n+\delta)^2}{R^2}+4m^2\sin^2\frac{\pi k}{2(N+1)}\quad
(n=\cdots, -1, 0, 1, \cdots, k=1,\cdots, N)\,.
\end{equation}
Therefore, one Dirac zero mode is found when $\delta=0$, since the matrix
$\tilde\mathbf{M}^2_n$ (for $\tilde{\psi}$) has no zero mode. If we replace
$m\rightarrow\frac{N+1}{\pi\tilde{R}}$ and taking the limit of
$N\rightarrow\infty$, the eigenvalues become
\begin{equation}
\frac{(n+\delta)^2}{R^2}+\frac{k^2}{\tilde{R}^2}\,.
\end{equation}
Six Dirac modes belong to the second-smallest eigenvalue in this limit, when
$\delta=0$ and $R=\tilde{R}$.
We thus recognized that the continuum limit of the model in this section
is a theory compactified on a cylinder.%
\footnote{In a crude sense, it can be said that the $S^1/Z_2\otimes C_{N}$ model
and the
$S^1\otimes P_{N+1}$ model are ``dual'' to each other.}

\subsection{$S^1/Z_2\otimes P_{N+1}$}

\begin{figure}[ht]
\centering
\includegraphics
{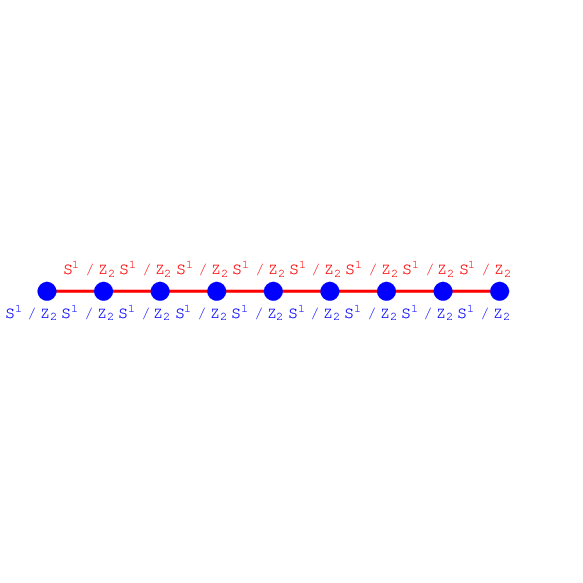}
\caption{Graphical representation of the $S^1/Z_2\otimes P_{N+1}$ model ($N=8$).}
\label{fig5_3}
\end{figure}

Next in this section, we consider the $S^1/Z_2\otimes P_{N+1}$ model
(Fig.~\ref{fig5_3}). As in the $S^1/Z_2\otimes C_{N}$ model, we assume that 
$\Psi$ and $\tilde{\Psi}$ have opposite parity conditions.
In this case, we find
\begin{eqnarray}
& &\int_0^{2\pi
R}\mathcal{L}_M[\Psi_p,
\tilde{\Psi}_p]dy=-m\sum_{p=1}^N\left[\overline{\tilde{\psi}}_{R,p,0}
(\psi_{L,p,0}-\psi_{L,p-1,0})+h.c.\right]\nonumber \\
& &\qquad-m\sum_{n=1}^\infty\sum_{p=1}^N\left[\overline{\tilde{\psi}}_{R,p,n}
(\psi_{L,p,n}-\psi_{L,p-1,n})+
\overline{\tilde{\psi}}_{L,p,n}
(\psi_{R,p,n}-\psi_{R,p-1,n})+h.c.\right]\,.
\end{eqnarray}

The mass-squared matrix $\mathcal{M}^2_n$ has the same structure as in the
$S^1\otimes P_{N+1}$ model in the previous subsection.
The eigenvalues of $\mathbf{M}^2_n$ are
\begin{equation}
\frac{n^2}{R^2}+4m^2\sin^2\frac{\pi k}{2(N+1)}\quad
(n=0, 1, \cdots, k=0, 1,\cdots, N)\,,
\end{equation}
while the eigenvalues of $\tilde\mathbf{M}^2_n$ are
\begin{equation}
\frac{n^2}{R^2}+4m^2\sin^2\frac{\pi k}{2(N+1)}\quad
(n=0, 1, \cdots, k=1,\cdots, N)\,.
\end{equation}

Thus, one chiral zero mode is found. If we replace
$m\rightarrow\frac{N+1}{\pi\tilde{R}}$ and taking the limit of
$N\rightarrow\infty$, the eigenvalues become
\begin{equation}
\frac{n^2}{R^2}+\frac{k^2}{\tilde{R}^2}\,.
\end{equation}
If further $R=\tilde{R}$, the number of the second-smallest Dirac eigenmodes is
three.

Judging from the geometric shape (an interval times an interval), the continuum
limit of the model in this subsection may be a theory compactified on a rectangle,
but the spectrum differs from the one for six-dimensional Dirac fermions on a
rectangle \cite{FHNST}.
This result may seem strange, but in the current model, the numbers of $\Psi_v$
and $\tilde{\Psi}_e$, that is, the numbers of vertices and edges, do not match, so
a discrepancy occurs in the number of eigenmodes.

\subsection{mimicking $T^2/Z_2$}

\begin{figure}[ht]
\centering
\includegraphics
{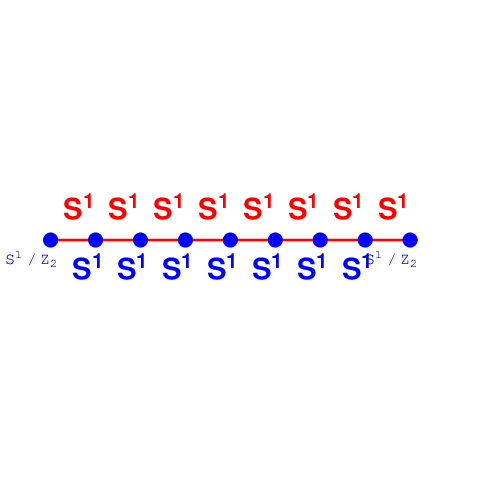}
\caption{Graphical representation of the $T^2/Z_2$-mimicker model.}
\label{fig5_4}
\end{figure}

We replace zeroth and $N$th $S^1$-compactified
theories in the $S^1\otimes P_{N+1}$ model with the 
$S^1/Z_2$-compactified theories 
(Fig.~\ref{fig5_4}). We anticipate that a ``mock'' orbifold with four fixed
points as $T^2/Z_2$ can be obtained by geometrically blocking a cylinder with two
line segments
$S^1/Z_2$.

The mass term in this case becomes
\begin{equation}
-\overline{\psi}_{R,0}\mathcal{M}_0\psi_{L,0}-\sum_{n=1}^\infty
\overline{\psi}_{R,n}\mathcal{M}_n\psi_{L,n}+h.c.\,,
\label{mf}
\end{equation}
where
$\psi_{L0}=
(\psi_{0,0},\dots,\psi_{N,0},\tilde{\psi}_{1,0},\dots,\tilde{\psi}_{N,0})^T_L$,
$\psi_{R0}=
(\psi_{1,0},\dots,\psi_{N-1,0},\tilde{\psi}_{1,0},\dots,\tilde{\psi}_{N,0})^T_R$,
and
$\psi_n=(\psi_{0,n},\dots,\psi_{N,n},\psi_{1,-n},\dots,\psi_{N-1,-n},\tilde{\psi}_{1,n},\dots,\tilde{\psi}_{N,n},
\tilde{\psi}_{1,-n},\dots,\tilde{\psi}_{N,-n})^T$, for $n\ge 1$.
The $(2N-1)\times(2N+1)$ matrix $\mathcal{M}_0$ is expressed by
\begin{equation}
\mathcal{M}_0=\left[\begin{array}{rr}
M_0 & {M''}_0\\
{M'_0} & \tilde{M}_0
\end{array}
\right]\,,
\end{equation}
and
\begin{equation}
{M}_0=\left[\begin{array}{rrrrr}
0&-i\delta_1/R & &&0\\
\vdots&&\ddots & &\vdots\\
0&&&-i\delta_{N-1}/R &0
\end{array}
\right]\,,\quad
\tilde{M}_0=\left[\begin{array}{rrr}
-i\tilde{\delta}_1/R & &\\
&\ddots & \\
&&-i\tilde{\delta}_N/R 
\end{array}
\right]\,,
\end{equation}
\begin{equation}
{M'}_0=m\left[\begin{array}{rrrrrr}
-1 & 1 & 0  &\cdots & 0 & 0\\
0  & -1 & ~1  &\cdots  &0 &0 \\
\vdots & \vdots & \vdots & \ddots & \vdots& \vdots  \\
0 & 0 & 0 & \cdots  & -1& ~1
\end{array}
\right]=-m E^T(P_{N+1})\,,
\end{equation}
\begin{equation}
{M''}_0=m\left[\begin{array}{rrrrrr}
1 & -1 & 0  &\cdots & 0 & 0\\
0  & 1 & -1  &\cdots  &0 &0 \\
\vdots & \vdots & \vdots & \ddots & \vdots& \vdots  \\
0 & 0 & 0 & \cdots  & 1& -1
\end{array}
\right]=m E^T(P_{N})\,.
\end{equation}

The $4N\times 4N$ matrix $\mathcal{M}_n$ is expressed by
\begin{equation}
\mathcal{M}_n=\left[\begin{array}{rr}
M_n & {{M}''}_n\\
{M'_n} & \tilde{M}_n
\end{array}
\right]\,,
\end{equation}
and
\begin{equation}
{M}_n=\left[\begin{array}{rrrrrrrr}
-i\frac{n}{R} &&&&&&&\\
&-i\frac{(n+\delta_1)}{R} &&&&&&\\
&&\ddots &&&&& \\
&&&-i\frac{(n+\delta_{N-1})}{R} &&&&\\
&&&&-i\frac{n}{R} &&&\\
&&&&&-i\frac{(-n+\delta_1)}{R} &&\\
&&&&&&\ddots & \\
&&&&&&&-i\frac{(-n+\delta_{N-1})}{R} \\
\end{array}
\right]\,,
\end{equation}
\begin{equation}
\tilde{M}_n=\left[\begin{array}{rrrrrrr}
-i\frac{(n+\tilde{\delta}_1)}{R} &&&&&&\\
&\ddots &&&&& \\
&&-i\frac{(n+\tilde{\delta}_{N})}{R} &&&&\\
&&&-i\frac{(-n+\tilde{\delta}_1)}{R} &&&\\
&&&&&\ddots & \\
&&&&&&-i\frac{(-n+\tilde{\delta}_{N})}{R} \\
\end{array}
\right]\,,
\end{equation}
\begin{equation}
{M'}_n=m\left[\begin{array}{rrrrrrrrrr}
-\frac{1}{\sqrt{2}} & 1 &   & &  & &&&&\\
  & -1 & ~1  &  & &&&&& \\
 &  & \ddots & \ddots & &  &&&& \\
&&  & -1 & 1  &  & &&& \\
 &  &  &   & -1& \frac{1}{\sqrt{2}}&&&&\\
-\frac{1}{\sqrt{2}}&&&&& & 1 &   & &  \\
&0&&&&  & -1 & 1  &  &  \\
&&&\ddots&& &  & \ddots & \ddots &   \\
&&&&0& &  &  & -1 & 1\\
&&&&&\frac{1}{\sqrt{2}} &  &  &   & -1
\end{array}
\right]\,,
\end{equation}
\begin{equation}
{{M}''}_n=m\left[\begin{array}{rrrrrrrrrr}
-\frac{1}{\sqrt{2}} &  & &  &\frac{1}{\sqrt{2}}&&&&&\\
1  & -1  &  & &&0&&&& \\
  & \ddots  &\ddots  & &&&&\ddots&& \\
 &   & 1  & -1& &&&&0&\\
 &   &   & \frac{1}{\sqrt{2}}& &&&&&-\frac{1}{\sqrt{2}}\\
&&& & 1 & -1  & & & &\\
&&&  &  & 1  & -1 &&& \\
&&& &  && \ddots & \ddots & & \\
&&& & & && 1 & -1 &  \\
&&& & & & & & 1 & -1
\end{array}
\right]\,.
\end{equation}

\begin{figure}[ht]
\centering
\includegraphics[width=6cm]
{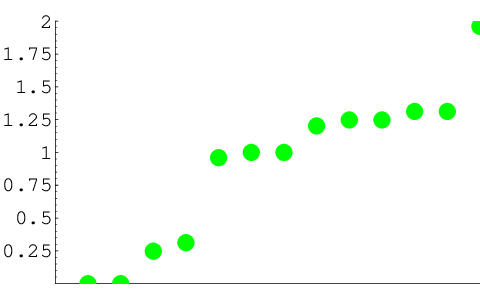}
\caption{The first several eigenvalues of the mass-squared matrix of
the ``mock'' orbifold model with vanishing gauge fields.
The parameters are chosen as: 
$m=\frac{N+1}{2\pi\tilde{R}}$, $R=\tilde{R}=1$, and $N=8$.}
\label{fig5_5}
\end{figure}

In Fig.~\ref{fig5_5}, the small eigenvalues are shown in the case without
background gauge field. The spectrum resembles the spectrum of the torus
compactification plus two massless chiral modes.

Next, we consider, as in the $S^1\otimes C_N$ model, the flux-like configuration
of the gauge fields:
\begin{equation}
\delta_p=\frac{qp}{N}\quad(p=1,\dots, N-1)\,,\quad
\tilde{\delta}_p=\frac{qp}{N}\quad(p=1,\dots, N)\,.
\label{5m}
\end{equation}

\begin{figure}[ht]
\centering
\includegraphics[width=6cm]
{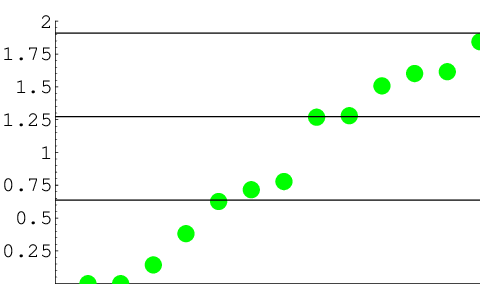}
\caption{The first several eigenvalues of the mass-squared matrix in the case that
$\delta_p$ and $\tilde{\delta}_p$ are given by (\ref{5m}) with
$q=1$, respectively. The other settings are,
$m=\frac{N+1}{2\pi\tilde{R}}$, $R=\tilde{R}=1$, and $N=8$.
The horizontal lines represent integer multiples of $\frac{2}{\pi}$.}
\label{fig5_6}
\end{figure}
\begin{figure}[ht]
\centering
\includegraphics[width=6cm]
{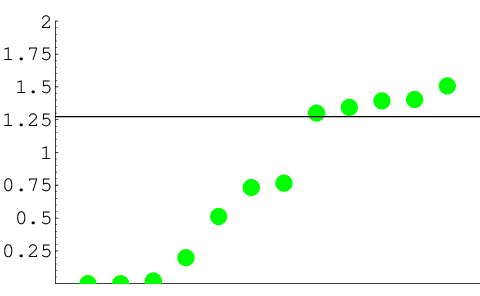}
\caption{The first several eigenvalues of the mass-squared matrix in the case that
$\delta_p$ and $\tilde{\delta}_p$ are given by (\ref{5m}) with
$q=2$, respectively. The other settings are,
$m=\frac{N+1}{2\pi\tilde{R}}$, $R=\tilde{R}=1$, and $N=8$.
The horizontal lines represent integer multiples of $\frac{4}{\pi}$.}
\label{fig5_7}
\end{figure}
\begin{figure}[ht]
\centering
\includegraphics[width=6cm]
{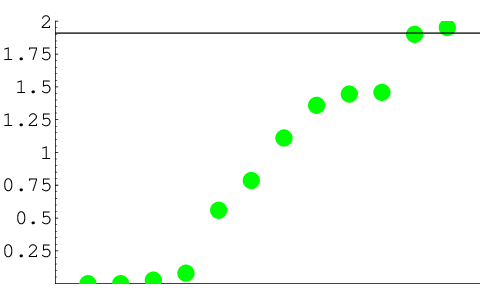}
\caption{The first several eigenvalues of the mass-squared matrix in the case that
$\delta_p$ and $\tilde{\delta}_p$ are given by (\ref{5m}) with
$q=3$, respectively. The other settings are,
$m=\frac{N+1}{2\pi\tilde{R}}$, $R=\tilde{R}=1$, and $N=8$.
The horizontal lines represent integer multiples of $\frac{6}{\pi}$.}
\label{fig5_8}
\end{figure}

\begin{figure}[ht]
\centering
\includegraphics[width=6cm]
{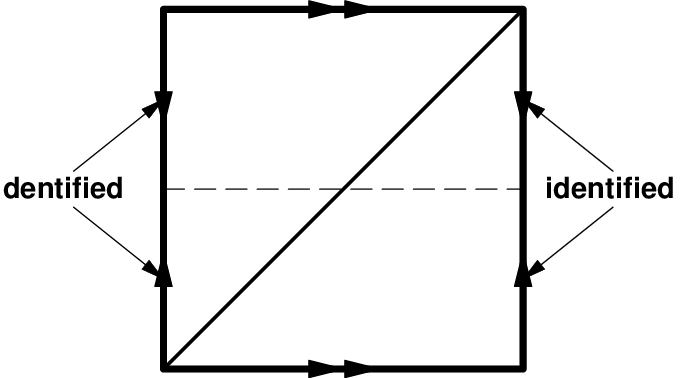}
\caption{We explain the factor two in the
spectral gap when $q=1$, for example. The $S^1$ structure is obtained by
identifying the top and bottom sides of the rectangle. The top and bottom
halves of the left and right sides are identified respectively, since we take
the quotient with $Z_2$. Meanwhile, the diagonal line in the figure
schematically represents
$\delta=R\langle A_y\rangle$. $\delta$ changes by unity between the left and
right ends, and then, the gauge fields on both ends are gauge equivalent. 
Since we take the quatient with
$Z_2$, the flux is the twice the value expected on a simple torus.}
\label{fig5_9}
\end{figure}

For $N=8$, the eigenvalues are shown in Figs.~\ref{fig5_6},
\ref{fig5_7}, and \ref{fig5_8},
when $q=1,2,3$, respectively.
It is known that the corresponding compactification yields
the eigenvalues located at integer multiples of $\frac{2q}{\pi R\tilde{R}}$.%
\footnote{The explanation for the factor two is illustrated in FIG.~\ref{fig5_9}.}
In our model, the relatively large gap appears at the same location, though the
degeneracies do not really match them. Thus, unfortunately, the continuum limit
of this model may not be a ``true ''
$T^2/Z_2$.
There are, in any case in the model, two chiral zero modes
which belong to
the zero eigenvalues of $\mathcal{M}_0$.

\section{Summary and discussion}
\label{conclusion}
In this paper, the mass spectra of light fermions are obtained in our
``half-deconstructed'' models. When compared with the spectrum for
manifolds, qualitative agreement was found when supposing the continuum limit.
Rather, from a particle physics perspective, it is an interesting toy model that
fermions with non-zero tiny mass appear in the models with graphs of finite size.
If we take the results into account, it can be realized that a very
small mass scale is generated in the partially deconstructed model, which is a
feature of this scheme. In that case, it is considered that the fermions in this
model are not genuine quarks or leptons. The scale which appears in the
model may have an indirect influence on the weak interaction scale or provides a
new scale of a model beyond the standard model.
For example, the scale may indicate that of the dark matter candidates, or 
may have the relevance to additional intermediate sectors which
interact only the fields at specific sites.
A clockwork mechanism \cite{CI,KR,GM,FPRT}%
\footnote{See also Refs.~\cite{CS,Sutherland}}
 has already been proposed as a
mechanism for generating such a small hierarchy,%
\footnote{Note that in clockwork mechanisms, clear and relatively large mass gap
(as in our model) generally do not appear.}
 so examining and comparing it
with our model, or constructing a fused model, will be an important future
challenge.

Since strict topological restrictions cannot be placed on the
configuration of our discrete fields, it will be necessary to study not only the
gauge field as a background field but also their dynamics. In other words, the
additional
bilinear term corresponding to the kinetic term of the gauge field may be
important in the whole Lagrangian of the model.

A natural extension of the model is one with varying the
compactification scale $R$ of each five-dimensional theory depending on the site.
It corresponds to warped manifolds such as the Randall--Sundrum theory \cite{RS}.
In that case, it will be necessary to consider mechanisms that give the
dynamics of such  scales. It is possible to randomly connect five-dimensional
theories with different compactifications with random configurations of gauge
fields or Wilson loops, but some kind of symmetry or guiding principle would be
required in such  cases.

The model discussed in this paper is characterized by not requiring the
introduction of new link fields corresponding to a component of a gauge
field on a discrete dimension. Although this is preferable from the standpoint of
simplicity, the feasibility of introducing such a link field in a modified
version of the model should be considered in the future, taking into account of
phenomenological models of particle physics.

In this paper, we focused on fermions and their modes of light mass eigenvalues,
but in future we should of course also discuss important research topics
such as theories involving degrees of freedom of gauge fields and scalar fields,
and models related to dark sectors. A model with supersymmetry can be constructed
relatively easily by replacing fields in the model here with superfields
\cite{KKS}. Therefore, we will be able to explore various possibilities for
supersymmetric models and their spectra, including semi-deconstructed models such
as those introduced here.

In this paper, we mainly introduced toy models with reference to the continuum
limits, but when a general graph is used, there is no counterpart of continuous
theory, and novel pseudo-extra-dimensional models become possible. The
deconstruction of higher dimensional manifolds (or orbifolds) is an interesting
extension. In the future, we would also like to consider full deconstruction,
that is, a four-dimensional theory that represents the extra dimensions by
graphs. In addition, it may be feasible that some other orbifold-like
compactification can be constructed based on the deconstruction  (for example,
\cite{SSK}) of a tilted torus (for a pedagogical introduction
\cite{Dienes}), so
 we wish to pursue its mathematical structure and applications.

\bibliographystyle{apsrev4-1}

\begin{thebibliography}{99}
\bibitem{KK} T.~Appelquist, M.~Chodos and P.~G.~O.~Freund,
\textit{Modern Kaluza--Klein theories},
Addison--Wisley, 1987.
\bibitem{ADD} N.~Arkani-Hamed, S.~Dimopoulos and G.~Dvali, 
``The hierarchy problem and new dimensions at a millimeter'',
Phys. Lett. \textbf{B429} (1998) 263. 
hep-ph/9803315.
\bibitem{ACG1}
N.~Arkani-Hamed, A.~G.~Cohen and H.~Georgi,
``(De)constructing dimensions'',
Phys. Rev. Lett. \textbf{86} (2001) 4757. 
hep-th/0104005.
\bibitem{HPW}
C.~T.~Hill, S.~Pokorski and J.~Wang,
``Gauge invariant effective Lagrangian for Kaluza--Klein modes'',
Phys. Rev. \textbf{D64} (2001) 105005.
hep-th/0104035.
\bibitem{Lane} K.~Lane,
``Case study in dimensional deconstruction'',
Phys. Rev. \textbf{D65} (2002) 115001.
hep-ph/0202093.
\bibitem{HL} C.~T.~Hill and A.~K.~Leibovich,
``Deconstructing 5-D QED'',
Phys. Rev. \textbf{D66} (2002) 016006.
hep-ph/0205057.
\bibitem{ACG2} N.~Arkani-Hamed, A.~G.~Cohen and H.~Georgi, 
``Electroweak symmetry breaking from dimensional deconstruction'',
Phys. Lett. \textbf{B513} (2001) 232.
hep-ph/0105239.
\bibitem{Schmaltz} M.~Schmaltz, 
``The simplest Little Higgs'',
JHEP \textbf{0408} (2004) 056.
hep-ph/0407143.
\bibitem{Hosotani1} Y.~Hosotani, 
``Dynamical mass generation by compact extra dimensions'',
Phys. Lett. \textbf{B126} (1983) 309.
\bibitem{Hosotani2} Y.~Hosotani, 
``Dynamics of nonintegrable phases and gauge symmetry breaking'',
Ann. Phys. (N.Y.) \textbf{190} (1989) 233.
\bibitem{HIL} H.~Hatanaka, T.~Inami and C.~S.~Lim, 
``The gauge hierarchy problem and higher dimensional gauge theories'',
Mod. Phys. Lett. \textbf{A13} (1998) 2601.
hep-th/9805067.
\bibitem{CKS} Y.~Cho, N.~Kan and K.~Shiraishi,
``Compactification in deconstructed gauge theory with topologically non-trivial
link fields'', Acta Phys. Polon. \textbf{35} (2004) 1597. 
hep-th/0306012.
\bibitem{Mohar1} B.~Mohar,
``{The Laplacian spectrum of graphs}'', in
\textit{Graph Theory, Combinatorics, and Applications},
Y. Alavi \textit{et al.} eds. (Wiley, New York, 1991), p.~871.
\bibitem{Mohar2} B.~Mohar,
``Laplace eigenvalues of graphs---a survey'',
Discrete Math. \textbf{109} (1992) 171. 
\bibitem{Mohar3} B.~Mohar,
``Some applications of Laplace eigenvalues of graphs'', in 
\textit{Graph Symmetry, Algebraic Methods, and Applications},
G. Hahn and G. Sabidussi eds.
(Kluwer, Dordrecht, 1997), p.~225.
\bibitem{Merris}
R.~Merris, 
``Laplacian matrices of graphs: a survey'', 
Linear Algebra Appl. \textbf{197} (1994) 143. 
\bibitem{PQ} See, for example, A.~Pomarol and M.~Quiros, 
``The standard model from extra dimensions'',
Phys. Lett. \textbf{B438} (1998) 255. 
hep-ph/9806263.
\bibitem{DS} M.~J.~Duncan and G.~C.~Segr\`e, 
``A simplified model for superstring compactification'', 
Phys. Lett. \textbf{B195} (1987) 36. 
\bibitem{DSW} M.~J.~Duncan, G.~C.~Segr\`e and J.~F.~Wheater, 
``Topological stability in higher dimensional theories'', 
Nucl. Phys. \textbf{B308} (1988) 509. 
\bibitem{TT} Y.~Tatsuta and A.~Tamiya, 
``(De)constructing magnetized dimensions'',
arXiv:1703.05263 [hep-th].
\bibitem{FHNST} Y.~Fujimoto, K.~Hasegawa, K.~Nishiwaki, M.~Sakamoto and
K.~Tatsumi, 
``6d Dirac fermion on a rectangle; scrutinizing boundary conditions, mode
functions and spectrum'', 
Nucl. Phys. \textbf{B922} (2017) 186. 
arXiv:1609.01413 [hep-th].
\bibitem{CI}
K.~Choi and S.~H.~Im,
``Realizing the relaxion from multiple axions and 
its UV completion with high scale supersymmetry'',
JHEP \textbf{1601} (2016) 149.
arXiv:1511.00132 [hep-ph].
\bibitem{KR}
D.~E.~Kaplan and R.~Rattazzi,
``Large field excursions and approximate
discrete symmetries from a clockwork axion'',
Phys. Rev. \textbf{D93} (2016) 085007.
arXiv:1511.01827 [hep-ph].
\bibitem{GM}
G.~F.~Giudice and M.~McCullough,
``A clockwork theory'',
JHEP \textbf{1702} (2017) 036.
arXiv:1610.07962 [hep-ph].
\bibitem{FPRT}
M.~Farina, D.~Pappadopulo, F.~Rompineve and A.~Tesi,
``The photo-philic QCD''
JHEP \textbf{1701} (2017) 095.
arXiv:1611.09855 [hep-ph].
\bibitem{CS} N.~Craig and D.~Sutherland,
``Exponential hierarchies from Anderson localization in theory space'',
Phys. Rev. Lett. \textbf{120} (2018) 221802.
arXiv:1710.01354 [hep-ph].
\bibitem{Sutherland} D.~Sutherland,
``Generating hierarchies with Anderson localization'',
Nucl. Part. Phys. Proc. \textbf{303--305} (2018) 59. 
\bibitem{RS} L.~Randall and R.~Sundrum, 
``A large mass hierarchy from a small extra dimension'',
Phys. Rev. Lett. \textbf{83} (1999) 3370.
hep-ph/9905221.
\bibitem{KKS} N.~Kan, K.~Kobayashi and K.~Shiraishi,
``Vortices and superfields on a graph'',
Phys. Rev. \textbf{D80} (2009) 045005.
arXiv:0901.1168 [hep-th].


\bibitem{SSK} K.~Shiraishi, K.~Sakamoto and  and N.~Kan, 
``Shape of deconstruction'',
J. Phys. \textbf{G29} (2003) 595. 
hep-ph/0209126.
\bibitem{Dienes} K.~R.~Dienes,
``Shape versus volume: Making large flat extra dimensions invisible'',
Phys. Rev. Lett. \textbf{88} (2002) 011601.
hep-ph/0108115.



\end{thebibliography}


\end{document}